\documentclass[aps, prb, twocolumn, reprint, superscriptaddress, floatfix, citeautoscript, longbibliography]{revtex4-1}

\pdfoutput=1

\usepackage[version=4]{mhchem}
\usepackage{siunitx}
\usepackage{graphicx}
\usepackage[T1]{fontenc}
\usepackage[utf8]{inputenc}
\usepackage[unicode,colorlinks=true,allcolors=blue]{hyperref}
\usepackage[para]{footmisc}
\usepackage{bm}
\usepackage[dvipsnames,table]{xcolor}
\usepackage{cleveref}

\usepackage[acronym]{glossaries}
\glsenableentrycount 

\newacronym{bnmr}{$\beta$-NMR}{$\beta$-detected NMR}
\newacronym{bnqr}{$\beta$-NQR}{$\beta$-detected nuclear quadrupole resonance}
\newacronym{bpp}{BPP}{Bloembergen-Purcell-Pound}
\newacronym{cw}{CW}{continuous wave}
\newacronym{efg}{EFG}{electric field gradient}
\newacronym{emim-ac}{\ce{EMIM-Ac}}{1-ethyl-3-methylimidazolium acetate}
\newacronym{fwhm}{FWHM}{full width at half maximum}
\newacronym{md}{MD}{molecular dynamics}
\newacronym{nmr}{NMR}{nuclear magnetic resonance}
\newacronym{pfg}{PFG}{pulsed field gradient}
\newacronym{rf}{RF}{radio frequency}
\newacronym{rtil}{RTIL}{room temperature ionic liquid}
\newacronym{slr}{SLR}{spin-lattice relaxation}
\newacronym{vft}{VFT}{Vogel-Fulcher-Tammann}
\newacronym{deme}{DEME}{N,N-diethyl-N-methyl-N-(2-methoxyethyl)\-ammonium}
\newacronym{tfsa}{TFSA}{bis\-(tri\-fluoro\-methane\-sulfonyl)\-amide}
\newacronym{fsa}{FSA}{bis\-(fluoro\-sulfonyl)\-amide }

\glsdisablehyper

\newcommand*{\elip}{\ce{^8Li^+}}
\newcommand*{\eli}{\ce{^8Li}}
\newcommand*{\lip}{\ce{Li^+}}
\newcommand*{\sli}{\ce{^7Li}}

\DeclareSIUnit\molar{\text{M}}
\DeclareSIUnit\ppm{\text{ppm}}
\DeclareSIUnit\gauss{\text{G}}
\DeclareSIUnit\hours{\text{h}}

\begin{document}
\title{The dynamics of liquid \glsdesc*{emim-ac} measured with implanted-ion \eli\ \glstext{bnmr}}

\author{Derek Fujimoto}
\email{fujimoto@phas.ubc.ca}
\affiliation{Department of Physics and Astronomy, University of British Columbia, Vancouver, BC V6T~1Z1, Canada}
\affiliation{Stewart Blusson Quantum Matter Institute, University of British Columbia, Vancouver, BC V6T~1Z4, Canada}

\author{Ryan M. L. McFadden}
\affiliation{Stewart Blusson Quantum Matter Institute, University of British Columbia, Vancouver, BC V6T~1Z4, Canada}
\affiliation{Department of Chemistry, University of British Columbia, Vancouver, BC V6T~1Z1, Canada}

\author{Martin H. Dehn}
\affiliation{Department of Physics and Astronomy, University of British Columbia, Vancouver, BC V6T~1Z1, Canada}
\affiliation{Stewart Blusson Quantum Matter Institute, University of British Columbia, Vancouver, BC V6T~1Z4, Canada}

\author{Yael Petel}
\affiliation{Department of Chemistry, University of British Columbia, Vancouver, BC V6T~1Z1, Canada}

\author{Aris Chatzichristos}
\affiliation{Department of Physics and Astronomy, University of British Columbia, Vancouver, BC V6T~1Z1, Canada}
\affiliation{Stewart Blusson Quantum Matter Institute, University of British Columbia, Vancouver, BC V6T~1Z4, Canada}

\author{Lars Hemmingsen}
\affiliation{Department of Chemistry, University of Copenhagen, 2100 København Ø, Denmark}

\author{Victoria L. Karner}
\affiliation{Stewart Blusson Quantum Matter Institute, University of British Columbia, Vancouver, BC V6T~1Z4, Canada}
\affiliation{Department of Chemistry, University of British Columbia, Vancouver, BC V6T~1Z1, Canada}

\author{Robert F. Kiefl}
\affiliation{Department of Physics and Astronomy, University of British Columbia, Vancouver, BC V6T~1Z1, Canada}
\affiliation{Stewart Blusson Quantum Matter Institute, University of British Columbia, Vancouver, BC V6T~1Z4, Canada}
\affiliation{TRIUMF, Vancouver, BC V6T~2A3, Canada}

\author{C. D. Philip Levy}
\affiliation{TRIUMF, Vancouver, BC V6T~2A3, Canada}

\author{Iain McKenzie}
\affiliation{TRIUMF, Vancouver, BC V6T~2A3, Canada}
\affiliation{Department of Chemistry, Simon Fraser University, Burnaby, BC V5A~1S6, Canada}

\author{Carl A. Michal}
\affiliation{Department of Physics and Astronomy, University of British Columbia, Vancouver, BC V6T~1Z1, Canada}

\author{Gerald D. Morris}
\affiliation{TRIUMF, Vancouver, BC V6T~2A3, Canada}

\author{Matthew R. Pearson}
\affiliation{TRIUMF, Vancouver, BC V6T~2A3, Canada}

\author{Daniel Szunyogh}
\affiliation{Department of Chemistry, University of Copenhagen, 2100 København Ø, Denmark}

\author{John O. Ticknor}
\affiliation{Stewart Blusson Quantum Matter Institute, University of British Columbia, Vancouver, BC V6T~1Z4, Canada}
\affiliation{Department of Chemistry, University of British Columbia, Vancouver, BC V6T~1Z1, Canada}

\author{Monika Stachura}
\email{mstachura@triumf.ca}
\affiliation{TRIUMF, Vancouver, BC V6T~2A3, Canada}

\author{W. Andrew MacFarlane}
\email{wam@chem.ubc.ca}
\affiliation{Stewart Blusson Quantum Matter Institute, University of British Columbia, Vancouver, BC V6T~1Z4, Canada}
\affiliation{Department of Chemistry, University of British Columbia, Vancouver, BC V6T~1Z1, Canada}
\affiliation{TRIUMF, Vancouver, BC V6T~2A3, Canada}

\date{\today}

\begin{abstract}
We demonstrate the application of implanted-ion \glsdesc*{bnmr} as a probe of ionic liquid molecular dynamics through the measurement of \eli\ \gls{slr} and resonance in \glsdesc*{emim-ac}. The motional narrowing of the resonance, and the local maxima in the \gls{slr} rate, $1/T_1$, imply a sensitivity to sub-nanosecond \lip\ solvation dynamics. From an analysis of $1/T_1$, we extract an activation energy ${E_A = \SI{74.8 \pm 1.5}{\milli\electronvolt}}$ and \glsdesc*{vft} constant ${T_{\mathrm{VFT}} = \SI{165.8 \pm 0.9}{\kelvin}}$, in agreement with the dynamic viscosity of the bulk solvent. Near the melting point, the lineshape is broad and intense, and the form of the relaxation is non-exponential, reflective of our sensitivity to heterogeneous dynamics near the glass transition. The depth resolution of this technique may later provide a unique means of studying nanoscale phenomena in ionic liquids.
\end{abstract}

\maketitle
\glsresetall	

\section{Introduction \label{sec:introduction}}

\Glspl{rtil} are a fascinating class of amorphous materials with many practical applications\cite{2015-Hayes-CR-115-6357,MacFarlane2014Rev}, such as lubrication in space applications and other low-pressure environments\cite{Haskins2016}. As in high temperature molten salts, strong Coulomb forces yield a liquid with significant \emph{structure}. Pair distribution functions from scattering experiments reveal an ion arrangement of alternating charges\cite{Tosi1993Rev,Murphy2015Rev,Bowron2010}, resulting in a large and strongly temperature dependent viscosity $\eta$. In contrast to simple salts, \glspl{rtil} consist of large, low-symmetry molecular ions and they remain liquid at ambient temperature. Many \glspl{rtil} are notoriously difficult to crystallize. Rather, they are easily supercooled, eventually freezing into a glassy state at the glass transition temperature $T_\mathrm{g}$ far below the thermodynamic melting point, $T_\mathrm{m}$\cite{Mudring2010Rev}.

A key feature of supercooled liquids and glasses is \emph{dynamic heterogeneity}\cite{Ediger2012Rev,Castner2010Rev,Sillescu1999Rev}. Distinct from homogeneous liquid or crystalline phases, the local \gls{md} exhibit fluctuations which are transient in both time and space. These non-trivial fluctuations are characterized by a growing dynamic correlation length, and are found to be stronger closer to the glassy phase\cite{Berthier2011d}. An understanding of dynamic heterogeneity may be central to a fundamental theoretical description of glass formation. 

With highly localized probes in the form of nuclear spins, \gls{nmr} is one of the few methods with the spatial and temporal resolution to quantify this heterogeneity and reveal its characteristics\cite{Kaplan1982,Khudozhitkov2018a,Sillescu1999Rev}. The degree of heterogeneity can be modelled by the ``stretching'' of an exponential nuclear \gls{slr}, ${\exp\left\{-[(\lambda t)^\beta]\right\}}$, where ${\lambda = 1/T_1}$ is the \gls{slr} rate and $\beta$ is the stretching exponent. Single exponential relaxation ($\beta=1$), corresponds to homogeneous dynamics, whereas $\beta<1$ describes a broad distribution of exponentials \cite{Lindsey1980}, the case where each probe nucleus relaxes at a different rate. The breadth of the distribution of rates is determined by $\beta$, with ${\beta=1}$ corresponding to a delta function. 

While stretched exponential relaxation is suggestive of dynamic heterogeneity, it is worth considering whether it instead results from a population which homogeneously relaxes in an intrinsically stretched manner.  To this point, \gls{md} simulations of a supercooled model binary liquid have shown $\beta$ to be independent of scale, at least down to a few hundred atoms\cite{Shang2019}. This implies that the stretching is intrinsic and homogeneous; however, the \gls{nmr} nuclei are each coupled to far fewer atoms, and are capable of identifying dynamical heterogeneity \cite{Kaplan1982,Khudozhitkov2018a}. This sensitivity is clearly demonstrated by 4D exchange \gls{nmr}, where subsets of nuclei in supercooled polyvinyl acetate were tracked by their local relaxation rate, revealing a broad distribution of relaxation times\cite{SchmidtRohr1991}. Furthermore, dynamical heterogeneities have been theoretically shown to be a prerequisite for stretched exponential relaxation in dynamically frustrated systems, such as supercooled liquids\cite{Simdyankin2003}. A reduction of $\beta$ below one is a signature of dynamic heterogeneity. 

\begin{figure}
	\centering
	\includegraphics[width=\columnwidth]{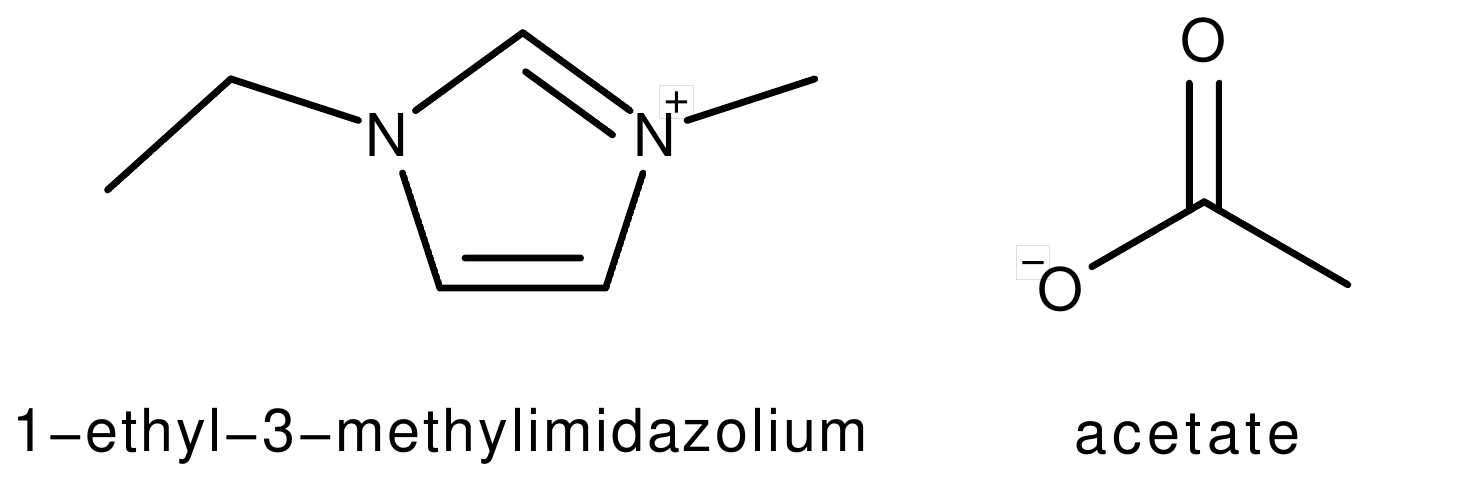} 
	\caption{Structure of the EMIM cation (\textit{left}) and Ac anion (\textit{right}).}
	\label{fig:emim-ac}
\end{figure}

Potential applications of the \gls{rtil} \gls{emim-ac}, with ions depicted in \Cref{fig:emim-ac}, have motivated detailed studies of its properties, including neutron scattering measurements of its liquid structure\cite{Bowron2010},  its bulk physical properties\cite{Bonhote1996,2011-Fendt-JCED-56-31,2011-Pinkert-PCCP-13-5136,2012-Pereiro-CC-48-3656,2016-Castro-JCED-61-2299,2009-Evlampieva-RJAC-82-666,2017-Zhang-JML-233-471,2015-Nazet-JCED-60-2400,2013-Araujo-JCT-57-1,2012-QuijadaMaldonado-JCT-51-51}, 
and its ability to dissolve cellulosic material\cite{2011-Freire-JCED-56-4813,2014-Castro-IECR-53-11850}. Here, we use implanted-ion \gls{bnmr} to study the development of dynamic heterogeneity and ionic mobility of implanted \elip\ in supercooled \gls{emim-ac}. The \gls{bnmr} signal is due to the anisotropic $\beta$-decay of a radioactive \gls{nmr} nucleus\cite{Heitjans2005,Ackermann1983,Heitjans1986}, similar to muon spin rotation. The probe in our case is the short-lived \eli, produced as a low-energy spin-polarized ion beam and implanted into the sample\cite{2015-MacFarlane-SSNMR-68-1}. At any time during the measurement, the \elip\ are present in the sample at ultratrace (\SI{e-13}{\molar}) concentration. Implanted-ion \gls{bnmr} has been developed primarily for studying solids, particularly thin films. It is not easily amenable to liquids, since the sample must be mounted in the beamline vacuum, yet the exceptionally low vapor pressure of \glspl{rtil} makes the present measurements feasible\cite{2018-Szunyogh-DT-47-14431}. 

We have measured the strong temperature dependence of the \gls{slr} ($1/T_1$) and resonance of \eli\ in \gls{emim-ac}. The relaxation shows a characteristic \gls{bpp} peak at \SI{298}{\kelvin}, coinciding with the emergence of dynamical heterogeneity, indicated by stretched exponential relaxation. Resonance measurements clearly demonstrate motional narrowing as the \gls{rtil} is heated out of the supercooled regime. Our findings show that \gls{bnmr} could provide a new way to study depth-resolved dynamics in \textit{thin films} of \glspl{rtil}\cite{Nishida2018film}.

\section{Experiment \label{sec:experiment}}

\gls{bnmr} experiments were performed at TRIUMF's ISAC facility in Vancouver, Canada. A highly polarized beam of \elip\ was implanted into the sample in the high-field \gls{bnmr} spectrometer with static field ${B_0 = \SI{6.55}{\tesla}}$~\cite{2014-Morris-HI-225-173,2004-Morris-PRL-93-157601}. The incident beam had a typical flux of \SI{\sim e6}{ions\per\second} over a beam spot \SI{\sim 2}{\milli\metre} in diameter. With a beam energy of \SI{19}{\kilo\electronvolt}, the average implantation depth was calculated by SRIM\cite{Ziegler2013} to be \SI{\sim200}{nm}, but solvent diffusion (see discussion) modifies this initial implantation profile significantly during the \eli\ lifetime. Spin-polarization of the \eli\ nucleus was achieved in-flight by collinear optical pumping with circularly polarized light, yielding a polarization of \SI{\sim 70}{\percent}\cite{Levy2002}. The \eli\ probe has nuclear spin ${I=2}$, gyromagnetic ratio ${\gamma = \SI{6.3016}{\mega\hertz\per\tesla}}$, nuclear electric quadrupole moment ${Q=\SI[retain-explicit-plus]{+32.6}{\milli b}}$, and radioactive lifetime ${\tau_{\beta}=\SI{1.21}{\second}}$. The nuclear spin-polarization of \eli\ is monitored through its anisotropic $\beta$-decay, where the observed \emph{asymmetry} of the $\beta$-decay is proportional to the average longitudinal nuclear spin-polarization\cite{2015-MacFarlane-SSNMR-68-1}. The proportionality factor is fixed and is determined by the $\beta$-decay properties of \eli\ and the detector geometry. The asymmetry of the $\beta$-decay was combined for two opposite polarizations of the \eli\ by alternately flipping the helicity of the pumping laser. This corrects for differences in detector rates and baseline\cite{Widdra1995,2015-MacFarlane-SSNMR-68-1}.

Similar to other quadrupolar (${I>1/2}$) nuclei in nonmagnetic materials, the strongest interaction between the \eli\ nuclear spin and its surroundings is typically the electric quadrupolar interaction, even when the time average of this interaction is zero. In \gls{emim-ac}, it is very likely that the spin relaxation is due primarily to fluctuations in the local \gls{efg} at the position of the \eli\ nucleus. The relaxation of a single ${I=2}$ nucleus is fundamentally bi-exponential, regardless of the functional form of the \gls{efg} spectral density, although the bi-exponential is not very evident in high fields where the faster exponential has a small relative amplitude\cite{Becker1982,Korblein1985}.

\Gls{slr} measurements used a pulsed \elip\ beam. The transient decay of spin-polarization was monitored both during and following the \SI{4}{\second} pulse, where the polarization approaches a steady-state value during the pulse, and relaxes to \num{\sim 0} afterwards. The effect is a pronounced kink at the end of the beam pulse, characteristic of \gls{bnmr} \gls{slr} data (Figure~\ref{fig:slr-spectra}). No \gls{rf} magnetic field is required for the \gls{slr} measurements, as the probe nuclei were implanted in a spin state already far from thermal equilibrium. Thus, it is typically faster and easier to measure \gls{slr} than to measure the resonance spectrum; however, this type of relaxation measurement has no spectral resolution, unlike to conventional \gls{nmr}, and reflects the spin relaxation of \emph{all} the \eli.

Resonances were acquired by stepping a \gls{cw} transverse \gls{rf} magnetic field slowly through the \eli\ Larmor frequency, with a continuous \elip\ beam. The spin of any on-resonance \eli\ is rapidly nutated by the \gls{rf} field, resulting in a loss in time-averaged asymmetry.

The sample consisted  of a \glsdesc{emim-ac} solution (Sigma-Aldrich). To avoid the response being dominated by trace-level \ce{Li}-trapping impurities, we introduced a stable isotope ``carrier'' (\ce{LiCl}) at low, but macroscopic concentration to saturate impurity \lip\ binding sites. Additional characterization of a similar solution, prepared in the same manner, can be found in the supplementary information of Ref.~\citenum{2018-Szunyogh-DT-47-14431}. The solution was kept in a dry-pumped rough vacuum for approximately \SI{12}{\hours} prior to the measurement. A \SI{\sim3}{\micro\liter} droplet was placed in a \SI{3}{\mm} diameter blank hole set \SI{0.5}{\mm} into a \SI{1}{\mm} thick aluminum plate. The Al plate was then bolted vertically into an ultrahigh vacuum (\num{e-10}~Torr) coldfinger liquid He cryostat and the temperature was varied from \SIrange{220}{315}{\kelvin}. The viscosity was sufficient to prevent the liquid from flowing out of the holder during the experiment. Sample mounting involved a few minutes exposure to air, followed by pumping for \SI{30}{\min} in the spectrometer's load lock.

Separately, we determined the self-diffusion coefficients of the \ce{LiCl} \gls{emim-ac} solution using conventional bi-polar \gls{pfg} \gls{nmr} with an in-house probe\cite{Michan2012} and spectrometer\cite{Michal2002} at \SI{8.4}{\tesla} and room temperature. A gradient pulse of ${\delta=\SI{3.2}{\ms}}$ was applied in varying strength, $g$, from \SIrange{50}{1200}{\gauss\per\cm}. The probe frequency was set to either \ce{^1H} or \sli, and the diffusion time $\Delta$ was varied between \SIrange{100}{450}{\ms}, according to the species diffusion rate. A delay of \SI{30}{\ms} allowed eddy currents to decay before acquisition. Diffusion coefficients were extracted by fitting the resulting Gaussian to the Stejkal-Tanner diffusion equation \cite{Stejskal1965}. 

\section{Results and Analysis \label{sec:results}}

\subsection{Relaxation \label{sec:results:relaxation}}

\begin{figure}
	\centering
	\includegraphics[width=\columnwidth]{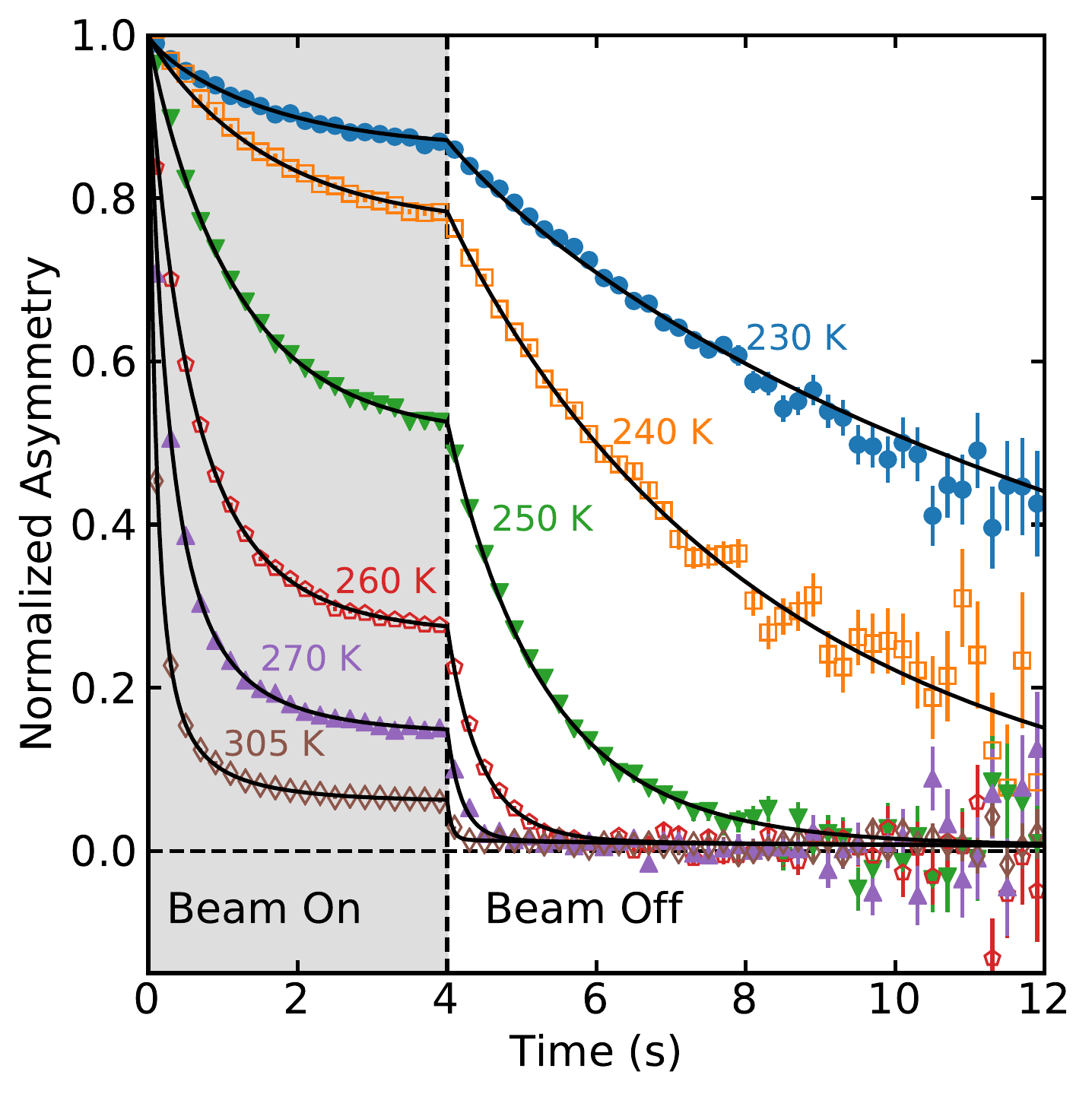}
	\caption{The $\beta$-decay asymmetry of \eli\ in \glstext{emim-ac}, with stretched exponential fits. The \Glstext{slr} is strongly temperature-dependent, and is well described by \Cref{eq:strexp} convoluted with the square beam pulse, as evidenced by $\tilde{\chi}^2_\mathrm{global} \approx 0.99$. The data have been binned by a factor of \num{20} for clarity. }
	\label{fig:slr-spectra}
\end{figure}

Typical \eli\ \gls{bnmr} \gls{slr} measurements are shown in \Cref{fig:slr-spectra}. Clearly, the relaxation shows a strong temperature dependence. At low temperatures it is slow, but its rate increases rapidly with temperature, revealing a maximum near room temperature. Besides the rate, the {\it form} of the relaxation also evolves with temperature. At low temperature it is highly non-exponential, but gradually crosses-over to nearly exponential at room temperature. For a \elip\ ion implanted at time $t^{\prime}$, the spin polarization $P(t,t^\prime)$ at time ${t>t^{\prime}}$ is well-described by a stretched exponential:
\begin{equation} \label{eq:strexp}
   P \left( t, t^\prime \right) = \exp \left\{ - \left[ \lambda \left( t-t^\prime \right) \right]^\beta \right\},
\end{equation}
where $t^\prime$ is integrated out as a result of convolution with the \SI{4}{\s} beam pulse\cite{2015-MacFarlane-PRB-92-064409}. A very small fraction, about \SI{2}{\percent}, of the \gls{slr} signal can be attributed to \elip\ stopping in the sample holder. While this background signal is nearly negligible, it is accounted for with an additive signal: ${P_b(t,t^\prime) = \exp\left\{-[\lambda_b(t-t^\prime)]^{0.5}\right\}}$, a phenomenological choice given the disorder in the \ce{Al} alloy. The relaxation rate, $\lambda_b$, was obtained from a separate calibration run at \SI{300}{\kelvin} and was assumed to follow the Korringa law: $\lambda_b \propto T$. 

The \gls{slr} time series at all $T$ were fit simultaneously with a common initial asymmetry. To find the global least-squares fit, we used C++ code leveraging the MINUIT~\cite{1975-James-CPC-10-343} minimization routines implemented within ROOT\cite{1997-Brun-NIMA-389-81}, accounting for the strongly time-dependent statistical uncertainties in the data. The fitting quality was excellent, with ${\tilde{\chi}^2_\mathrm{global} \approx 0.99}$. 

\begin{figure}
	\centering
	\includegraphics[width=\columnwidth]{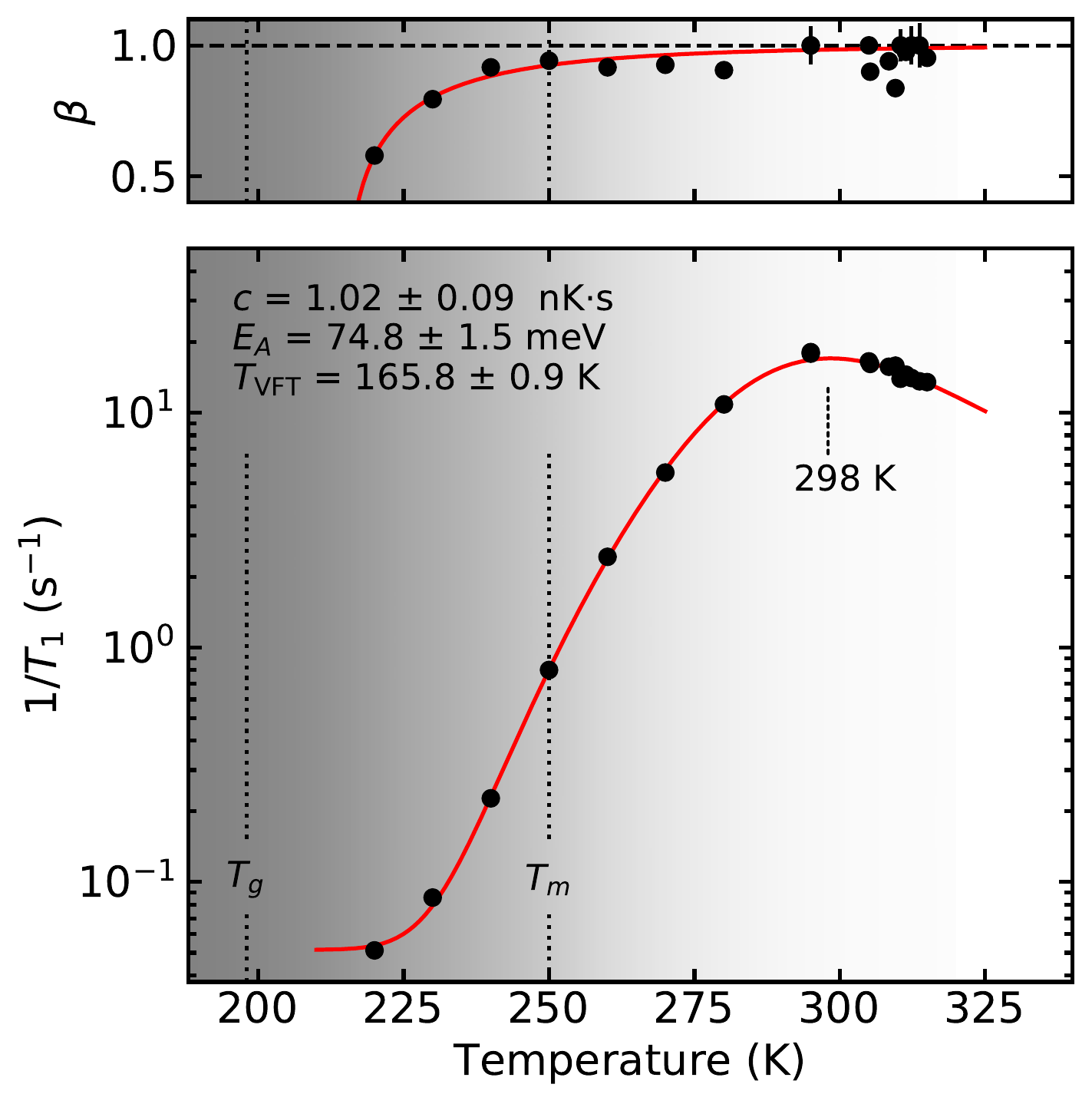}
	\caption{The stretched exponential parameters from fits to the \glstext{slr} in \gls{emim-ac} (refer to \Cref{fig:slr-spectra} for fit curves). For the rate ($1/T_1$), the line denotes a fit using \Cref{eq:model,eq:j,eq:tau}, as detailed in the text. For the stretching exponent ($\beta$), the line is a guide for the eye. Both are highly temperature dependent, showing a monotonic increase until the $1/T_1$ \glstext{bpp}\cite{1948-Bloembergen-PR-73-679} peak at \SI{298}{\kelvin}, above which $\beta\approx1$.  The shaded region indicates the approximate temperature range of supercooling between $T_\mathrm{g}$ and $T_\mathrm{m}$. } 
	\label{fig:slr-fits}
\end{figure}

As shown in \Cref{fig:slr-fits}, the change in $1/T_1$ over the measured \SI{~100}{\kelvin} range is remarkable, varying over 3 orders of magnitude. These changes coincide with the relaxation converging to monoexponentiality with increasing temperature, as evidenced by $\beta \rightarrow 1$ (upper panel). The temperature dependence of $1/T_1$ is, however, not monotonic; the rate is clearly maximized at room temperature, corresponding to a \gls{bpp} peak\cite{1948-Bloembergen-PR-73-679}. At this temperature, the characteristic fluctuation rate of the dynamics responsible for the \gls{slr} ($\tau_c^{-1}$) matches the probe's Larmor frequency (${\omega_0=\gamma B_0}$), i.e., ${\tau_c\omega_0 \approx 1}$. The \gls{slr} due to a fluctuating \gls{efg} can be described by the following simple model\cite{Abragam1962}:
\begin{equation} \label{eq:model}
   \frac{1}{T_1} = a \left[ J_1(\omega_0) + 4J_2(2\omega_0) \right] + b,
\end{equation}
where $a$ is a coupling constant related to the strength of the \gls{efg}, $b$ is a small phenomenological temperature-independent relaxation rate important at low $T$\cite{1991-Heitjans-JNCS-131-1053}, and the $J_{n}$ are the $n$-quantum \gls{nmr} fluctuation spectral density functions. If the local dynamics relax exponentially, the spectral density has the Debye (Lorentzian) form:
\begin{equation} \label{eq:j}
   J_n(n\omega) = \frac{2 \tau_c}{1 + \left(n \omega \tau_c \right)^2},
\end{equation}
where $\tau_c$ is the (exponential) correlation time. 
\begin{figure}
	\centering
	\includegraphics[width=\columnwidth]{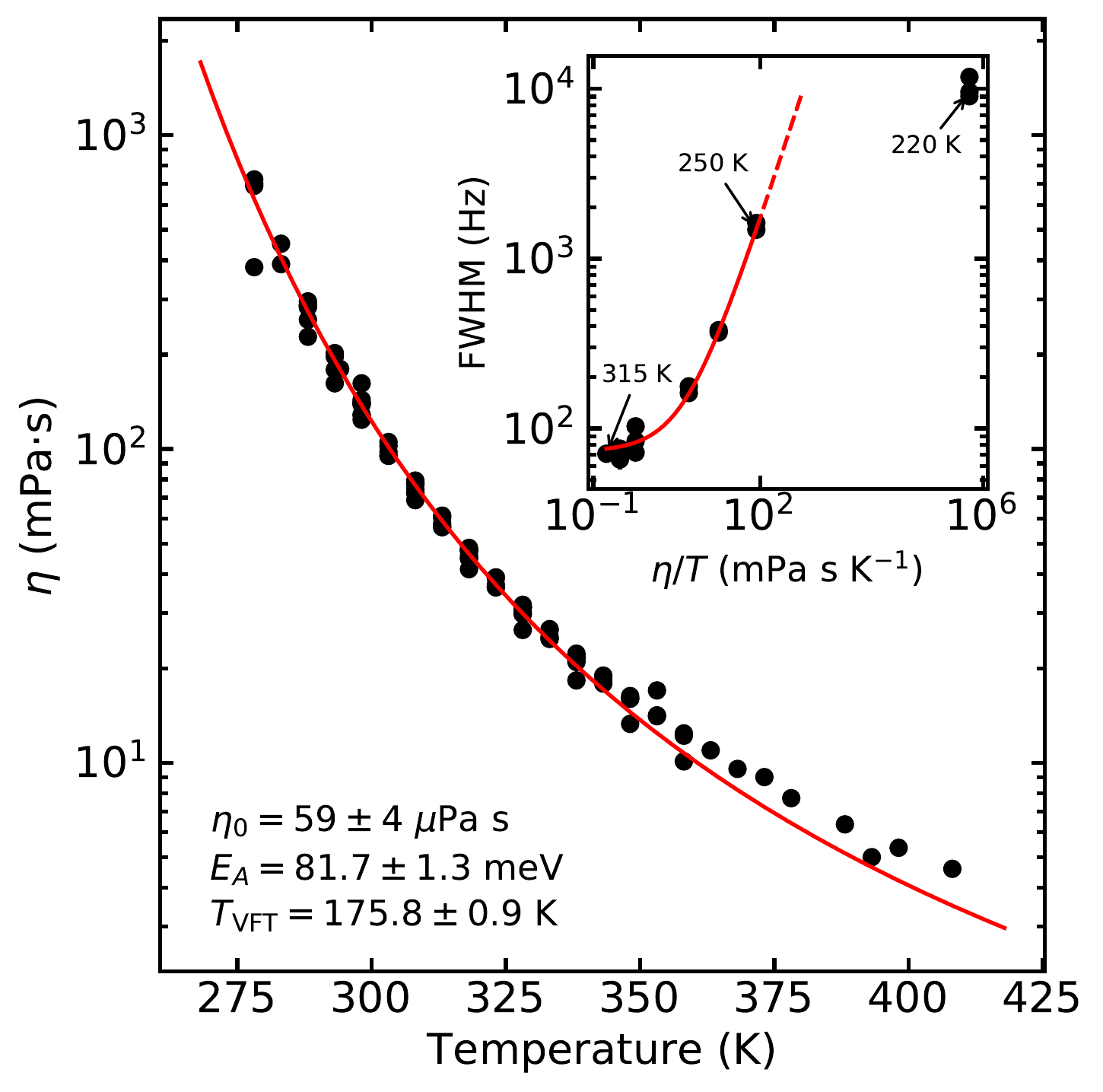}
	\caption{The dynamic viscosity ($\eta$) from the literature\cite{Bonhote1996, 2009-Evlampieva-RJAC-82-666, 2011-Fendt-JCED-56-31, 2011-Freire-JCED-56-4813, 2011-Pinkert-PCCP-13-5136, 2012-Pereiro-CC-48-3656, 2012-QuijadaMaldonado-JCT-51-51, 2013-Araujo-JCT-57-1, 2014-Castro-IECR-53-11850, 2015-Nazet-JCED-60-2400, 2016-Castro-JCED-61-2299, 2017-Zhang-JML-233-471}, fitted with a \glstext{vft} model. \glstext{emim-ac} is a fragile glass former, as evidenced by a super-Arrhenius $\eta$. \emph{Inset}: the resonance linewidth as a function of $\eta(T)/T$, with linear fit for $T\ge \SI{250}{\kelvin}$. This linear scaling is expected from the Stokes-Einstein relation.}
	\label{fig:viscosity}
\end{figure}

Local fluctuations may be related to other macroscopic properties of the liquid such as the viscosity. Using values from the literature~\cite{Bonhote1996, 2009-Evlampieva-RJAC-82-666, 2011-Fendt-JCED-56-31, 2011-Freire-JCED-56-4813, 2011-Pinkert-PCCP-13-5136, 2012-Pereiro-CC-48-3656, 2012-QuijadaMaldonado-JCT-51-51, 2013-Araujo-JCT-57-1, 2014-Castro-IECR-53-11850, 2015-Nazet-JCED-60-2400, 2016-Castro-JCED-61-2299, 2017-Zhang-JML-233-471}, \Cref{fig:viscosity} shows that the dynamic viscosity ($\eta$) of \gls{emim-ac} is non-Arrhenius, characteristic of a fragile glass-former, and can be described with the phenomenological \gls{vft} model. The inset shows that the linewidth is proportional to $\eta/T$, consistent with the Stokes-Einstein relation (Equation 48 of \citet{1948-Bloembergen-PR-73-679}). We then assume that $\tau_c$ is proportional to $\eta/T$:
\begin{equation} \label{eq:tau}
   \tau_c=\frac{c}{T}\exp\left[ \frac{E_A}{k_B \left(T-T_\mathrm{VFT}\right)} \right],
\end{equation} 
where $c$ is a prefactor, $E_A$ is the activation energy, $k_B$ is the Boltzmann constant, $T$ is the absolute temperature, and $T_\mathrm{VFT}$ is a constant. Together, \Cref{eq:model,eq:j,eq:tau} encapsulate the temperature and frequency dependence of the \eli\ $1/T_1$ in the supercooled ionic liquid. A fit of this model to the data is shown in \Cref{fig:slr-fits}, and parameter values can be found in \Cref{tbl:fitpar}. The correlation times from \SIrange{220}{315}{\kelvin} are on the order of nanoseconds. The choice of \Cref{eq:j} assumes that the $\beta <1$ stretching arises from a population of exponential relaxing environments with a broad distribution of $\tau_c$. As mentioned, this assumption is likely good for the \eli\ \gls{bnmr} probe; especially since the basic local relaxation of \eli\ due to quadrupolar coupling is not intrinsically stretched, independent of the dynamical fluctuation spectrum\cite{Becker1982}. Under this construction, the departure from ${\beta=1}$ in the supercooled regime is consistent with the emergence of dynamical heterogeneity.
\begin{table}
\bgroup
\arrayrulecolor{Gray!70}
\begin{tabular}{ll|ll}
&& $\bm{1/T_1(T)}$ & $\bm{\eta(T)}$ \\ \hline
$c$ 			&(\si{n\K\cdot\s})			& \num{1.02\pm0.09}&\\
$\eta_0$ 		&(\si{\micro\Pa\cdot \s})	& & \num{59\pm4}\\
$E_A$ 			&(\si{\meV})				& \num{74.8\pm1.5} 			& \num{81.7\pm1.3}\\
$T_\mathrm{VFT}$&(\si{\K})					& \num{165.8\pm0.9} 		& \num{175.8\pm0.9}\\
$a$				&(\si{\s^{-2}})				& \num{1.550\pm0.006e9}&\\
$b$				&(\si{\per\s})				& \num{5.13\pm0.07e-2}&
\end{tabular}
\egroup
\caption{Fit results for $1/T_1$ (\gls{bnmr}) and $\eta$ (literature). Parameters are defined in \Cref{eq:model,eq:tau}, substituting $c/T\leftrightarrow\eta_0$ in the latter. Corresponding curves are shown in \Cref{fig:slr-fits,fig:viscosity}.}
\label{tbl:fitpar}
\end{table}

\subsection{Resonance \label{sec:results:resonance}}

\begin{figure}
	\centering
	\includegraphics[width=\columnwidth]{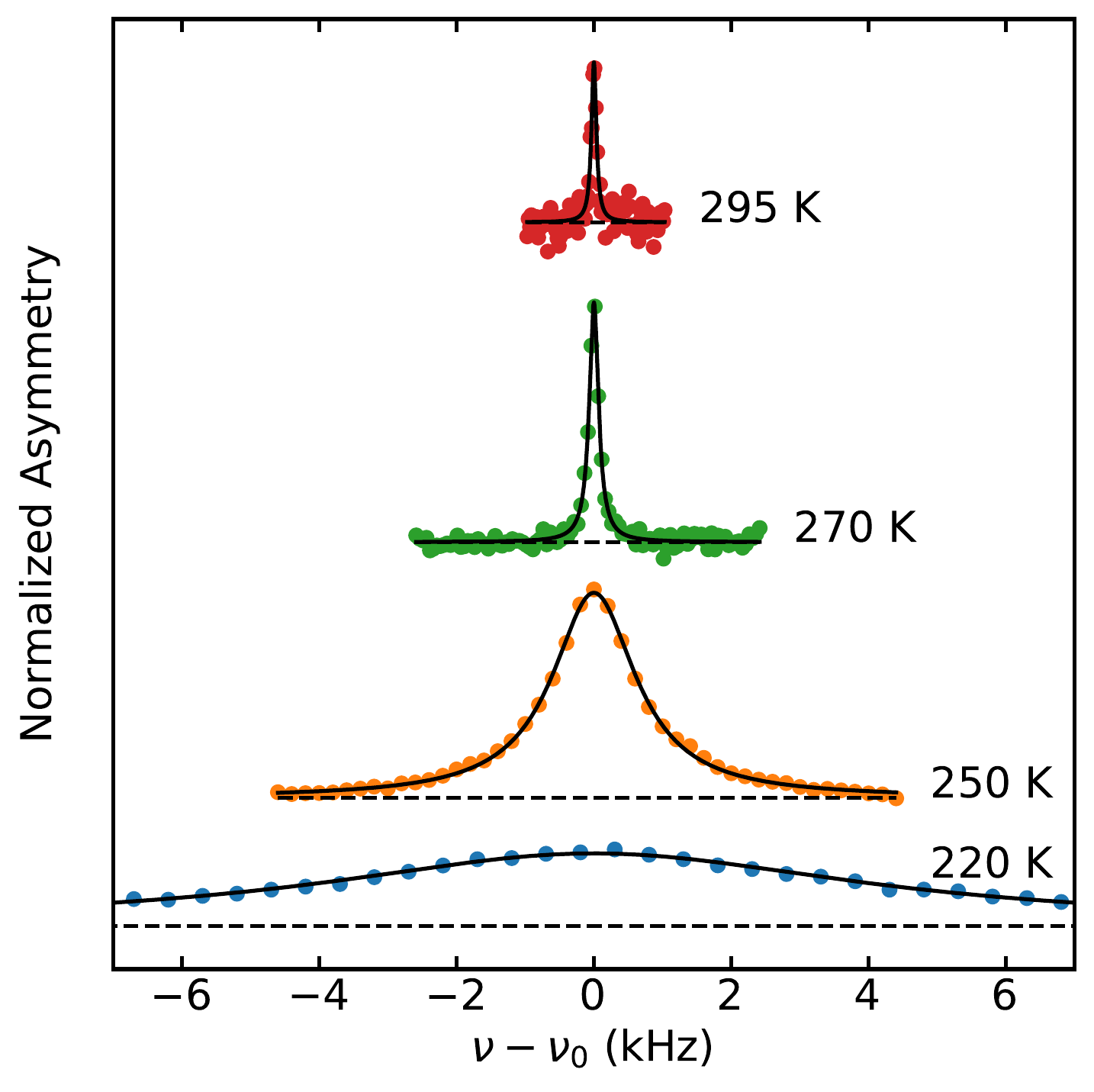}
	\caption{The \eli\ resonance in \glstext{emim-ac}, shifted by the Larmor frequency ($\nu_0\approx\SI{41.27}{\MHz}$), with Lorentzian fit. The line narrows and increases in height as the temperature is raised, with a peak in the latter near \SI{260}{\kelvin} (see \Cref{fig:resonance-fits}). The vertical scale is the same for all spectra, which have been offset for clarity. Spectra are inverted for consistency with the presentation in conventional \gls{nmr}.}
	\label{fig:resonance-spectra}
\end{figure}

Typical \eli\ resonances are shown in \Cref{fig:resonance-spectra}. Similar to the \gls{slr}, they show a strong temperature dependence. At low $T$, the resonance is broad with a typical solid-state linewidth on the order of \SI{10}{\kilo\hertz}. The lack of resolved quadrupolar splitting reflects the absence of a single well-defined \gls{efg}; the width likely represents an inhomogeneous distribution of static, or partially averaged, \glspl{efg} giving a broad ``powder pattern'' lineshape convoluted with the \gls{cw} \gls{nmr} excitation, a Lorentzian of width ${\gamma B_1}$, where $B_1 \approx\SI{0.1}{\gauss}$. This inhomogeneous quadrupolar broadening is qualitatively consistent with the heterogeneity in the dynamics implied by the stretched exponential relaxation. 

\begin{figure}
	\centering
	\includegraphics[width=\columnwidth]{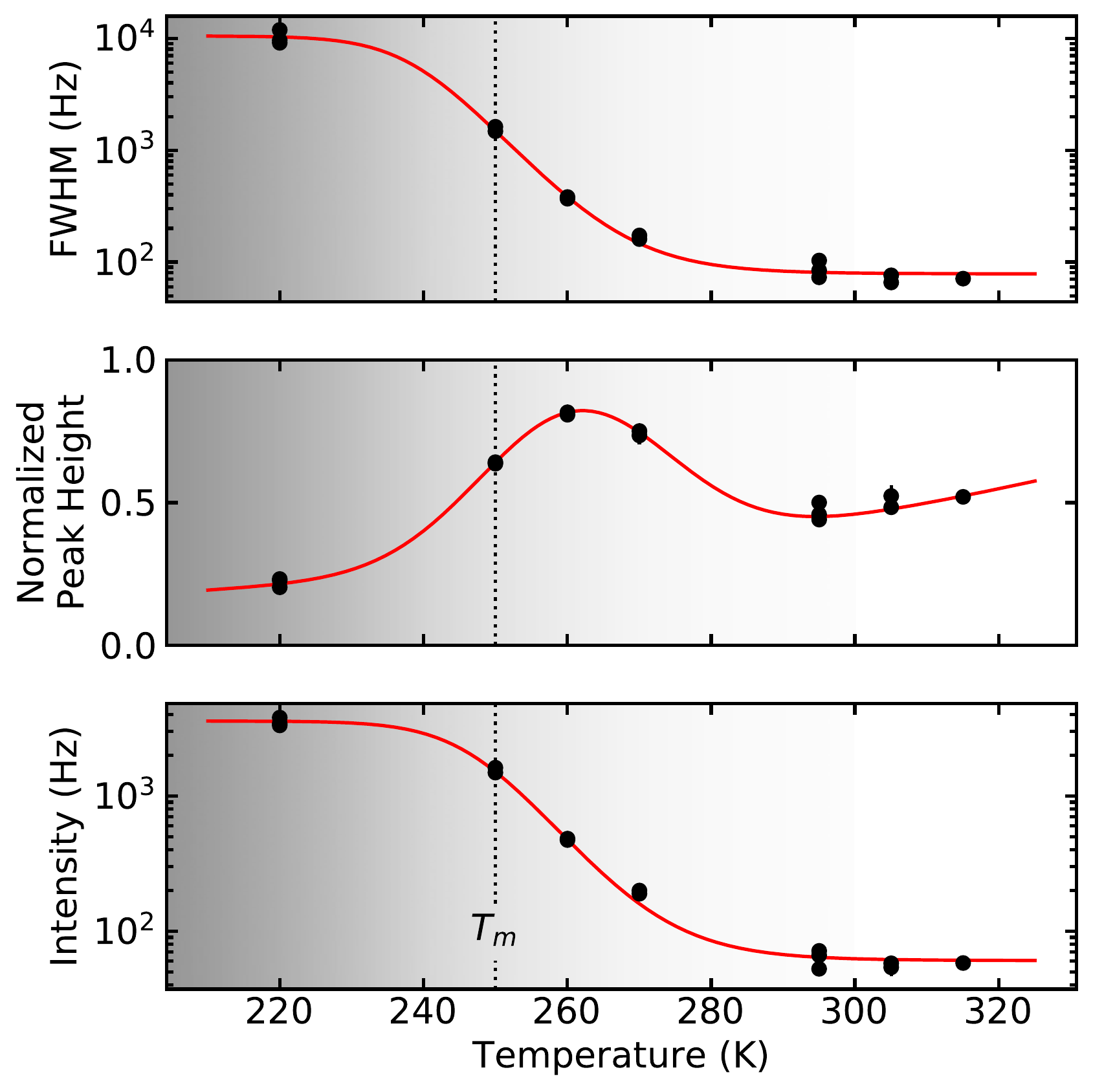}
	\caption{The Lorentzian fit parameters of the \eli\ resonance in \gls{emim-ac}, illustrated in \Cref{fig:resonance-spectra}, with lines to guide the eye. Narrowing of the line suggests an onset of solvent molecular motion above the melting point. The corresponding drop in intensity (area of the normalized spectra), and the non-monotonic peak height suggests inhomogeneous broadening at low temperature, and slow spectral dynamics occurring on the scale of \SI{1}{\s}, the integration time at each frequency. Shading indicates the supercooled region between $T_\mathrm{m}$ and $T_\mathrm{g}$ (off scale).}
	\label{fig:resonance-fits}
\end{figure}

The resonances are well-described by a simple Lorentzian. The baseline (time-integrated) asymmetry is also strongly temperature dependent due to the temperature dependence of $1/T_1$. The shift of the resonance relative to a single crystal of \ce{MgO} (our conventional frequency standard) is about \SI{-9}{\ppm}, but a slow drift of the magnetic field prevents a more accurate determination or a reliable measurement of any slight $T$ dependence. The other fit parameters extracted from this analysis; the linewidth, peak height, and intensity (area of normalized spectra); are shown in \Cref{fig:resonance-fits}.

As anticipated from the most striking features in \Cref{fig:resonance-spectra}, the linewidth and peak height evolve considerably with temperature. Note that the peak height in \Cref{fig:resonance-fits} is measured from the baseline, and is normalized to be in units of the baseline, accounting for changes in the \gls{slr}. Reduction in the linewidth by several orders of magnitude is compatible with motional narrowing, where rapid molecular motion averages out static inhomogeneous broadening. Saturation of the narrowing by room temperature
\footnote{The high temperature linewidth (\SI{\sim1.6}{\ppm}) is compatible with the limit imposed by the homogeneity of the magnet at its center (\SI{\sim10}{\ppm} over a cubic centimeter).}
 with an onset far below the $1/T_1$ maximum is consistent with the \gls{bpp} interpretation of the \gls{slr} peak~\cite{1948-Bloembergen-PR-73-679}.

\section{Discussion \label{sec:discussion}}

Mediated by a strong Coulomb interaction, \glspl{rtil} are known to contain a significant amount of structure. One might expect pairing of anions and cations, but calculations based on a simplified ion interaction model suggest that such pairs are short-lived\cite{Lee2015dilute}. Dielectric relaxation experiments confirm this, placing a \SI{100}{\pico\second} upper bound on their lifetime, rendering them a poor description of the average ionic structure\cite{Daguenet2006}. Rather, the arrangement can be described as two interpenetrating ionic networks. As revealed by neutron scattering \cite{Tosi1993Rev,Murphy2015Rev,Bowron2010}, each network forms cages about the other that are highly anisotropic due to the tendency for \ce{EMIM} rings to stack \cite{Bowron2010}. In fragile glass formers, such as \gls{emim-ac}, \gls{md} simulations indicate that the motion of the caged ion and the center of mass motion of the cage are correlated\cite{Habasaki2015}. 
Presumably, in our case, the small \elip\ cation is coordinated by several acetates and a similar correlation will exist for the \elip\ in the absence of independent long-range diffusion.

Naturally, the motion of the surrounding ionic solvent cage will cause the local \gls{efg} to fluctuate, and a strong temperature dependence is reasonable since these same fluctuations have a role in determining the strongly temperature dependent viscosity $\eta(T)$ shown in \Cref{fig:viscosity}. While a direct relation between the specific motions sensed by \eli\ and the bulk $\eta$ is complex and unclear\cite{1987-Akitt-MNNMR-7-189}, one may anticipate a consistency between their kinetics should a single mechanism govern both. The similarity of both $E_A$ and $T_\mathrm{VFT}$ with those found from the viscosity of the pure \gls{emim-ac} suggests that this is the case and provides further justification for the choice of \Cref{eq:tau}. 


The inset of \Cref{fig:viscosity} shows that motional narrowing causes the resonance linewidths to scale as ${\eta/T}$ in the liquid state above $T_\mathrm{g}$, a situation also observed in \glstext{deme}-\glstext{tfsa} [\glstext{deme} = \glsdesc{deme}; and \glstext{tfsa} = \glsdesc{tfsa}] with solute \sli\ \gls{nmr}\cite{Shirai2011LiNMR}. That this relationship holds for \eli\ is surprising; our \gls{bnmr} signal is due to the dynamics of a population of implanted local probes, for which solvent self-diffusion and probe tracer-diffusion are not differentiated, whereas the viscosity is a bulk property. If \elip\ is diffusing, it implies that the diffusion is controlled by the solvent dynamics. In the limiting case of a solid, interstitial diffusion can be fast, yet the viscosity infinite, and the decoupling of diffusion and the host viscosity is self-evident. Many \glspl{rtil} violate the Stokes-Einstein relation that linearly relates self-diffusivity $D$ to ${T/\eta}$, and its violation at low $T$ in the inset of \Cref{fig:viscosity} shows that ionic diffusion in supercooled \glspl{rtil} may contain some of the character expected from a solid. At \SI{295}{\K} however, our \sli\ \cgls{pfg} \gls{nmr} in \gls{emim-ac} with \SI{30}{\m\molar} \ce{LiCl} shows that the diffusion is not significantly larger than the solvent (${D_{\ce{Li}}=\SI{3.46\pm0.11 e-10}{\m^2\s^{-1}}}$ vs ${D_{\ce{H}}=\SI{3.61\pm0.07 e-10}{\m^2\s^{-1}}}$), demonstrating that the \eli\ is primarily sensing the mobility of its surrounding solvent cage. 

Relatively little is known about \lip\ as a solute in \gls{emim-ac}, compared to other imidazolium-based \glspl{rtil}, which have been explored as electrolytes for lithium-ion batteries\cite{Garcia2004batt,Shirai2011LiNMR}. Their properties should be qualitatively comparable, but the details certainly differ as both anion size and shape play a role in the diffusivity\cite{Tokuda2005PFGNMR}. Shown to compare favorably with implanted-ion \gls{bnmr}\cite{2018-Szunyogh-DT-47-14431}, conventional \gls{nmr} can provide a comparison to some closely related \glspl{rtil}: \ce{EMIM-TFSA} and \ce{EMIM-FSA} [\glstext{fsa} = \glsdesc{fsa}]. In both cases, the diffusion of \sli\ was similar to that of the solvent ions\cite{Hayamizu2011}. Differences in the tracer diffusion are reflected in the activation barrier for \sli\ hopping: \SI{222\pm6}{\milli\electronvolt} and \SI{187\pm2}{\milli\electronvolt}, respectively\cite{Hayamizu2011}. This correlates well with anion molecular weight, \SI{280}{\g\per\mol} and \SI{180}{\g\per\mol}, and with the barrier we report for \eli: \SI{74.8\pm1.5}{\milli\electronvolt} for acetate of \SI{59}{\g\per\mol}. This further emphasizes the probe sensitivity to the solvent dynamics. 

The motional narrowing immediately apparent in \Cref{fig:resonance-fits} is analogous to conventional pulsed \gls{rf} \gls{nmr}, but the use of \gls{cw} \gls{rf} modifies the detailed description significantly. While the details are beyond the scope of this work, and will be clarified at a later date, we now give a qualitative description. In the slow fluctuation regime, the line is \emph{broadened} relative to the static limit at $T=0$ due to slow spectral dynamics occurring over the second-long integration time at each RF frequency. Both the peak height and the intensity (area of the normalized curve) are increased through the resulting double counting of spins at multiple \gls{rf} frequencies. In the fast fluctuation limit, the time spent with a given local environment is small and the RF is relatively ineffective at nutating the spins. Unlike the slow fluctuation limit, transverse coherence is now needed to destroy polarization. Coherence is maintained only in a small range about the Larmor frequency, narrowing as the fluctuation rate increases. The intensity (area) is also reduced from the preservation of off-resonance polarization, but the peak height is unaffected. 

The local maximum in the peak height is explicable from the small slow relaxing background. When the \gls{rf} is applied on resonance, the signal from the sample is eliminated and the asymmetry is independent of the \gls{slr}. Increasing the \gls{slr} will, however, reduce the off-resonance asymmetry and results in a reduction in the fraction of destroyed polarization. This competes with the increase in peak height from motional narrowing and produces the local maximum in \Cref{fig:resonance-fits}. 

The development of dynamic heterogeneity at the nanosecond timescale ($\omega_0^{-1}$) is demonstrated by the stretched exponential \gls{slr}, as shown in \Cref{fig:slr-fits}. Concurrently, the line broadening shows that this heterogeneity reaches down to the static timescale. There are no definitive measurements of the melting point of \gls{emim-ac}, since it has not yet been crystallized, but $T_\mathrm{m}$ is no larger than \SI{250}{K}\cite{Sun2009}. In contrast, a calorimetric glass transition has been observed at about \SI{\sim198}{K}\cite{Guan2011}. Thus, the dynamic inhomogeneity develops in a range of $T$ that corresponds well to the region of supercooling, indicated by the shading in \Cref{fig:slr-fits}. Stretched exponential relaxation, reflecting dynamic heterogeneity, is a well-known feature of \gls{nmr} in disordered solids\cite{Bohmer2000Rev,Schnauss1990}. In some cases, diffusive spin dynamics, driven by mutual spin flips of identical near-neighbor nuclei, can act to wash out such heterogeneity. Such spin diffusion may be quenched by static inhomogeneities that render the nuclei non-resonant with their neighbors\cite{Bernier1973}. However, a unique feature of \gls{bnmr} is that spin diffusion is absent: even in homogeneous systems, the probe isotope is always distinct (as an \gls{nmr} species) from the stable host isotopes, and the \gls{bnmr} nuclei are, themselves, always isolated from one another. In the absence of spin diffusion, on quite general grounds, it has been shown\cite{1991-Heitjans-JNCS-131-1053,Stockmann1984} that the stretching exponent $\beta$ should be 0.5. Our data in \Cref{fig:slr-fits} appear to be approaching this value at the lowest temperatures. While stretched exponential relaxation is very likely a consequence of microscopic inhomogeneity, unequivocal confirmation requires more sophisticated measurements such as spectral resolution of the SLR and reduced 4D-\gls{nmr}\cite{SchmidtRohr1991}.

Based on the non-Arrhenius behaviour of $\eta(T)$, \gls{emim-ac} is a reasonably fragile glass former, comparable to toluene which has been studied in some detail using \ce{^2H} \gls{nmr}\cite{Hinze1996Toluene,Bohmer2000Rev}, providing us with a useful point of comparison to a nonionic liquid. Like \eli, $^2$H should exhibit primarily quadrupolar relaxation. Toluene is supercooled between its melting point \SI{178}{\kelvin} and glass transition \SI{\sim 117}{\kelvin}, though it shows stretched exponential relaxation only below about $1.1~T_\mathrm{g}$, considerably deeper into the supercooled regime than in our case, with an onset near $1.25~T_\mathrm{g}$, likely due to the stronger tendency to order in the ionic liquid. 

The closest analogue to our experiment is, perhaps, an early (neutron activated) \eli\ \gls{bnmr} study in \ce{LiCl.7D2O}\cite{1991-Heitjans-JNCS-131-1053,Faber1989}. There, the observed temperature dependence of the \gls{slr} is qualitatively similar (see Figure~9 of Ref.~\citenum{1991-Heitjans-JNCS-131-1053}): at low temperatures, the relaxation is nearly temperature independent, followed by a rapid increase above the glass transition, leading eventually to the \gls{bpp} peak at higher temperatures. This behaviour was interpreted as the onset of molecular motion above \SI{\sim 80}{\kelvin}, whose characteristic correlation times reflect the diffusion and orientational fluctuations in \ce{D2O}. This is consistent with the picture outlined here, although in our more limited temperature range the relaxation can be ascribed to a single dynamical process.

At present, there are few examples of \eli\ \gls{bnmr} in organic materials, as this application is in its infancy. Nevertheless, several trends from these early investigations have emerged, which serve as an important point of comparison. From an initial survey of organic polymers~\cite{2014-McGee-JPCS-551-012039}, it was remarked that resonances were generally broad and unshifted, with little or no temperature dependence. In contrast, the SLR was typically fast and independent of the proton density, implying a quadrupolar mechanism caused by the \gls{md} of the host atoms. These dynamics turned out to be strongly depth dependent, increasing on approach to a free surface\cite{2015-McKenzie-SM-11-1755} or buried interface\cite{2018-McKenzie-SM-14-7324}. In addition to dynamics of the polymer backbone, certain structures admitted \lip\ diffusion\cite{2014-McKenzie-JACS-136-7833}, whose mobility was found to depend on the ionicity of the anion of the dissolved \ce{Li} salt\cite{2017-McKenzie-JCP-146-244903}. A few small molecular glasses have also been investigated, where the relaxation is similarly fast\cite{2018-Karner-JPSCP-21-011022}.

Common to all of these studies is the non-exponential decay of the \eli\ spin-polarization, which is well described by a stretched exponential. In these disordered materials, the ``stretched'' behavior is compatible with the interpretation of a distribution of local environments, leading to an inhomogeneous \gls{slr}. Due to their high $T_\mathrm{g}$, the dynamics did not homogenize below the spectrometer's maximum temperature of \SI{\sim315}{\K}, unlike \gls{emim-ac}. This work is an important first example where the liquid state is attainable to a degree where we recover simple exponential \gls{slr}, accompanied by motional narrowing and a \gls{bpp} peak.

\section{Conclusion \label{sec:conclusion}}

We report the first measurements of \eli\ \gls{bnmr} in the ionic liquid \glsdesc{emim-ac}. Our results demonstrate that the quadrupolar interaction does not hinder our ability to follow the \gls{bnmr} signal through both the liquid and glassy state. We observed clear motional narrowing as the temperature is raised, accompanied by enhanced spin-lattice relaxation, whose rate is maximized at room temperature. From an analysis of the temperature dependent \gls{slr} rate, we extract an activation energy and \gls{vft} constant for the solvation dynamics, which are in relatively good agreement with the dynamic viscosity of (bulk) \gls{emim-ac}. At low temperatures near $T_\mathrm{m}$, the resonance is broad and intense, reflective of our sensitivity to slow heterogeneous dynamics near the glass transition. In this temperature range, the form of the relaxation is well-described by a stretched exponential, again indicative of dynamic heterogeneity. These findings suggest that \eli\ \gls{bnmr} is a good probe of both solvation dynamics and their heterogeneity. The depth resolution of ion-implanted \gls{bnmr} may provide a unique means of studying nanoscale phenomena in ionic liquids, such as ion behaviour at the liquid-vacuum interface or the dependence of diffusivity on film thickness~\cite{2018-Maruyama-ACSN-12-10509}.

\begin{acknowledgments}
The authors thank R.~Abasalti, D.~J.~Arseneau, S.~Daviel, B.~Hitti, and D.~Vyas for their excellent technical support.
This work was supported by NSERC Discovery grants to RFK, CAM, and WAM;
AC and RMLM acknowledge the support of their NSERC CREATE IsoSiM Fellowships;
MHD, DF, VLK, and JOT acknowledge the support of their SBQMI QuEST Fellowships;
LH thanks the Danish Council for Independent Research | Natural Sciences for financial support.
\end{acknowledgments}

\bibliography{references/references,references/emim-ac-viscosity,references/bnmr-liquids,references/ionic-liquids,references/emim-ac,references/andrew,references/derek,references/michal}

\end{document}